\documentstyle[multicol,aps,prl,psfig]{revtex}
 
\renewcommand{\narrowtext}{\begin{multicols}{2} \global\columnwidth20.5pc}
\renewcommand{\widetext}{\end{multicols} \global\columnwidth42.5pc}
\begin{document}

\title{Effects of Trilinear Term in  Softly Broken $N=1$ Supersymmetric QCD}

\author{M. Chaichian$^1$, W.F. Chen$^2$ and T. Kobayashi$^1$}

\address{$^1$ Department of Physics, 
University of Helsinki and Helsinki Institute of Physics\\
P.O. Box 9 FIN-00014 Helsinki, Finland\\
$^2$ Winnipeg Institute for Theoretical Physics and Department of Physics,\\ 
University of Winnipeg, Winnipeg, Manitoba, Canada R3B 2E9} 
\maketitle

\begin{abstract}
Softly broken dual magnetic theory of $N=1$ supersymmetric 
$SU(N_c)$ QCD with $N_f$ flavours is investigated with the inclusion of
trilinear coupling term of scalar fields in the case of $N_f> N_c+1$. 
It is found that the the trilinear soft supersymmetric breaking term
greatly change the phase and the vacuum structure. 
\end{abstract}



\narrowtext

There has been a big progress in understanding strongly coupled
$N=1$ supersymmetric Yang-Mills theory in the last few years
\cite{ref1,ref2}.
A complete phase diagram of the theory, the particle spectrum
and the dynamical phenomenon in each phase have been quantitively
or qualitatively figured out. In particular, it was found that
$N=1$ supersymmetric QCD with gauge group $SU(N_c)$ and $N_f$ flavour
quarks possesses a ``conformal window" $3N_c/2 <N_f <3N_c$ in the
infrared region of the theory when $N_f>N_c+1$, where the theory
can be described by a physically equivalent $N=1$ supersymmetric 
$SU(N_f-N_c)$ theory with $N_f$ flavours and $N_f^2$ singlets   but with the
strong and weak coupling be exchanged and vice versa. This is exactly
the realization of the old Montonen-Olive non-Abelian electric-magnetic
duality conjecture \cite{ref3} in $N=1$ supersymmetric gauge theory, which 
exists only in $N=4$ \cite{ref4} and has an analogue in 
low-energy $N=2$ \cite{ref5} supersymmetric gauge theories.      

It is natural to extend these non-perturbative analysis
to the ordinary QCD since its non-perturbative aspect is not clear
yet. To do this, the first step is breaking supersusymmetry. The 
most convenient breaking method is the
the introduction of superparticle mass terms such as squarks
and gaugino\cite{ref6,ref7,ref8}. However, when $N_f>N_c+1$ 
the dual magnetic theory
has an additional flavour interaction superpotential. 
A trilinear soft supersymmetry 
breaking term composed of the scalar fields is thus allowed. In this talk
we shall emphasize the effects of this trilinear term in determining
the phase structure and the vacuum structure of the soft broken
$N=1$ dual QCD\cite{ref9}. 



$N=1$ supersymmetric QCD with gauge group $SU(N_c)$ and $N_f$ flavour
quark chiral supermultiples $Q_{ir}$, $\widetilde{Q}_{ir}$
$i=1,{\cdots},N_f$, $r=1,{\cdots},N_c$, has an anomaly free global symmetry
\begin{eqnarray}
SU(N_f)_L{\times}SU(N_f)_R{\times}U(1)_B {\times}U(1)_R
\label{eq1}
\end{eqnarray}
At the low-energy the quark supermultiplets are confined, the dynamical
degrees of freedom are the chiral supermultiplets, meson $M_{ij}$ and 
baryons $B^{[i_1{\cdots}i_{N_c}]}$ and $\widetilde{B}^{[i_1{\cdots}i_{N_c}]}$. 
The low-energy effective theory still 
possesses the global symmetry (\ref{eq1}). When $N_f>N_c+1$, the dynamics
of $M$, $B$ and $\widetilde{B}$ can be described  $N=1$ supersymmetric 
$SU(N_f-N_c)$ gauge theory with $N_f$ flavours of quark chiral supermultiplets
 $q_{i\widetilde{r}}$, $\widetilde{q}_{i\widetilde{r}}$ and colour 
singet chiral fields ${\cal M}_{ij}$,
$i,j=1,{\cdots},N_f$, $\tilde{r}=1,{\cdots},N_f-N_c$, and an additional
flavour interaction superpotential
\begin{eqnarray}
W=\widetilde{q}^{i\tilde{r}}{\cal M}_{ij}q^{jr}
\label{eq2}
\end{eqnarray}
This conjecture is supported by 't Hooft anomaly matching.

The most important property in supersymmetric field theory is 
non-renormalization theorem, which determine the superpotential
must be a holomorphic function of the chiral superfield. The holomorphy 
of superpotential and global symmetry (\ref{eq1}) as well as the
instanton calculation can give a series of exact result of
low-energy supersymmetric QCD. Since supersymmetric QCD is very sensitive
to the relative numbers of colour and flavours, so we list
the non-perturbative dynamical phenomena according to the different
ranges of the colour and flavour numbers.

When $N_f<N_c$ there will dynamically generate a
superpotential, which eliminates all
the supersymmetry vacua.
In the case $N_f=N_c$,  the non-perturbative quantum correction modify
the classical moduli space constrained by $\det M-B\widetilde{B}=0$
as the quantum one, $\det M-B\widetilde{B}=\Lambda^{2N_c}$.
Depending on the vacuum choice in the moduli space,
the theory can present various dynamical patterns.
For examples, the vacuum
$M^i_j=\Lambda^2\delta^i_j$, $B=\widetilde{B}=0$
leads to the chiral symmetry breaking and confinement; while the other
vacuum choice $M^i_j=0$, $B=-\widetilde{B}=\Lambda^{N_c}$ makes 
chiral symmetry unbroken  and the baryon number violation.
In the case $N_f=N_c+1$,
the quantum moduli space is the same as classical moduli space. 
Consequently, the low-energy theory present confinement 
but no chiral symmetry breaking. 
When $N_f>N_c+1$,
if $N_f > 3N_c$, the theory at high energy level is not 
asymptotically free and hence the low-energy theory in a free
electric phase, the coupling constant behaves as 
$\alpha (R)\sim {1}/{\ln (R\Lambda)}$;
The more interesting is the range 
$3N_c/2<N_f<3N_c$, here the theory can have an non-trivial IR fixed point, 
at which the low-energy theory becomes an interacting conformal field theory.
Thus it is called Seiberg's conformal window. The $SU(N_f-N_c)$
theory describes the same physics as the high energy $SU(N_c)$ QCD,
but with the strong and weak coupling exchanged and vice versa
This is called the non-Abelian electro-magnetic duality..

%
When $N_f<N_c$ the quadratic soft supersymmetry breaking term 
at low-energy can be written out near the origin of the moduli space
\cite{ref6}
\begin{eqnarray}
{\cal L}_{\rm sb}&=&\int d^4\theta \left[B_TM_Q \mbox{Tr}(M^{\dagger}M)
+B_BM_Q\left(B^{\dagger}B\right.\right.\nonumber\\
&+&\left.\left.\widetilde{B}^{\dagger}\widetilde{B} \right)\right]
-\left[\int d^2\theta M_g \langle W^\alpha W_\alpha)+h.c. \right]
\end{eqnarray}
In the case $N_f{\geq}N_c+1$, 
the composite superfields are equivalently replaced by the dual
magnetic quarks and the soft breaking Lagrangian is\cite{ref6,ref7}
\begin{eqnarray}
{\cal L}_{\rm sb}=B_M m^2_M\mbox{Tr}(\phi_M^{\dagger}\phi_M)+B_qm^2_q
(\phi_q^{\dagger}\phi_q+\phi_{\widetilde{q}}^{\dagger}\phi_{\widetilde{q}})
\label{eq5}
\end{eqnarray}
In the decoupling limit, $m_g, m_Q{\rightarrow}\infty$, 
the features of the ordinary QCD are expected. 

The non-perturbative dynamical features in this soft broken 
supersymmetric QCD had been analyzed\cite{ref6}. The results show that
when $N_f < N_c$, the standard $SU_V(N_f){\times}U(1)_B$ QCD vacuum  
can arise, while in the case 
 $N_f = N_c$, there emerges an
exotic vacua with chiral symmetry
and spontaneously breaking of the baryon number symmetry, i.e. the vacuum 
is invariant $SU_L(N_f){\times}SU_R(N_f){\times}U(1)_R$.
When $N_f > N_c$,
There is a vacuum state with unbroken chiral symmetry,
but it is interesting that the non-Abelian electric-magnetic duality 
persists in the presence of soft supersymmetry breaking.  


When $N_f {\geq}N_c+1$, in addition to (\ref{eq5}),
the trilinear term 
\begin{eqnarray}
{\cal L}^{\prime}_{SB}=h\phi_{qi}\phi^i_{Mj}\phi^j_{\widetilde{q}}
\label{eq6}
\end{eqnarray}
can also make the supersymmetry softly broken
due to the superpotential (\ref{eq2}), where $h$ is the
trilinear coupling constant.
With the inclusion of (\ref{eq6}) the scalar potential becomes\cite{ref9}
\begin{eqnarray}
&&V(\phi_q, \phi_{\widetilde{q}},\phi_M)=
\frac{1}{k_T}\mbox{Tr}\left(\phi_q\phi^{\dagger}_q
\phi^{\dagger}_{\widetilde{q}}\phi_{\widetilde{q}}\right)\nonumber\\
&&+\frac{1}{k_T}\mbox{Tr}\left(\phi_q\phi_M\phi_M^{\dagger}\phi_q^{\dagger}
+{\phi}^{\dagger}_{\widetilde{q}}\phi_M^{\dagger}\phi_M
{\phi}_{\widetilde{q}}\right)
\nonumber\\
&&+\frac{\widetilde{g}^2}{2}\left(\mbox{Tr}\phi_q^{\dagger}\widetilde{T}^a
\phi_q-\mbox{Tr}\phi_{\widetilde{q}}\widetilde{T}^a
\phi_{\widetilde{q}}^{\dagger}\right)^2\nonumber\\
&& +m_q^2\mbox{Tr}(\phi_q^{\dagger}\phi_q)
+m_{\widetilde{q}}^2
\mbox{Tr}(\phi_{\widetilde{q}}^{\dagger}\phi_{\widetilde{q}})
+m_M^2\mbox{Tr}(\phi_M^{\dagger}\phi_M)
\nonumber\\
&&-\left(h\mbox{Tr}\phi_{qi}\phi^i_{Mj}\phi_{\widetilde{q}}^j+h.c.\right)
\label{eq7}
\end{eqnarray}
where $\widetilde{T}^a$ are the 
generators of magnetic gauge group $SU(N_f-N_c)$,
$\widetilde{g}$ is the magnetic gauge coupling constant. 
The phase (or vacuum) structure can be revealed by analyzing the minima
of (\ref{eq7}). 
With assumption that $h$ is real,
the minimum of potential can be obtained along diagonal
direction
\begin{eqnarray}
&&\phi_{qi}^r=\left\{\begin{array}{ll}\phi_{q(i)}\delta^{r}_{~i} & 
i=1,{\cdots}, N_f-N_c\\
 0 & \mbox{otherwise}\end{array}\right.\nonumber\\
&&\phi_{\widetilde{q}i}^r=\left\{\begin{array}{ll}
\phi_{\widetilde{q}(i)}\delta^{r}_{~i} &
i=1, {\cdots}, N_f-N_c\\
 0 & \mbox{otherwise}\end{array}\right.\nonumber\\
&&\phi_{Mj}^i=\left\{\begin{array}{l}\phi_{M(i)}\delta^{i}_{~j}, 
~~i,j=1,{\cdots}, N_f-N_c\\
0, ~~\mbox{otherwise}\end{array}\right.
\end{eqnarray} 
The analysis shows that in the direction 
$\phi_{q(i)}=q$ and $\phi_{\widetilde{q(i)}}=0$ (or 
$\phi_{q(i)}=0$ and $\phi_{\widetilde{q}(i)}=q$), 
 the vacuum expectation value
$\langle \phi^i_{Mj}\rangle =0$.
If $m_q^2 >0$ (or  $m_{\widetilde{q}}^2 >0$),  the scalar potential
has the minimum $V=0$ at $q=0$, thus theory is in 
chiral symmetric phase. 
However, if $m_q^2 <0$ (or  $m_{\widetilde{q}}^2 <0$),
the scalar potential unbounded from below and 
the theory becomes unphysical. In the flat direction of $D$-term, 
$\phi_{q(i)}=\phi_{\widetilde{q}(i)}=X_i$, 
if soft breaking parameters satisfy 
$h^2{\geq}{2}/{k_q}\left(m_q^2+m_{\widetilde{q}}^2\right)$,
then we find 
\begin{eqnarray}
&&\frac{h-\sqrt{h^2-2(m_q^2+m_{\widetilde{q}}^2)/k_q}}{2/k_q}{\leq}\langle
\phi_{M(i)}\rangle \nonumber\\ 
&& {\leq}\frac{h+\sqrt{h^2-2(m_q^2+m_{\widetilde{q}}^2)/k_q}}{2/k_q}
\label{eq8p12}
\end{eqnarray}
Thus chiral symmetry broken phase arises and the baryon number 
violation occurs.
Furthermore, depending on the ratio 
$\rho{\equiv}2m_M^2/(m_q^2+m_{\widetilde{q}}^2) k_q/k_M$,
the phase structure in 
$D$-flat directions presents various patterns. For examples, 
in the phase
diagram labeled by $(h^2, (m_q^2+m_{\widetilde{q}}^2)/2)$, when $\rho=1$,
the theory only has one chiral symmetry 
broken phase and one unbroken phase iff all $m_q^2$, 
$m_{\widetilde{q}}^2$ and $m_M^2$ are positive, whereas when $\rho=20$
theory has two unbroken phases and two chiral symmetry broken phases.
If $m_q^2$, $m_{\widetilde{q}}^2$ and $m_M^2$ 
are negative, scalar potential is unbounded from below and 
becomes unphysical.
Furthermore, in chiral symmetric phase, the $SU(N_f)^3$ and $SU(N_f)^2U(1)_B$ 
't Hooft anomalies match. In the broken phase, the two softly broken dual
theories also present same anomaly structure. This seems to suggest
Seiberg's duality remains after SUSY breaking,
 even in the chiral symmetry broken phase.

In the case $N_f=N_c+1$, the flavour symmetry (\ref{eq1}) allows a
trilinear soft breaking term $h'\phi_{Bi}\phi_{Mj}^i\phi_{\widetilde{B}}^j$.
With the combination of the quadratic term and the terms from
the effective potential
${\cal W}=(B_i M^i_j\widetilde{B}^j -\det M)/{\Lambda^{2N_c-1}}$,
the whole scalar potential is\cite{ref9}
\begin{eqnarray}
V &=& \lambda^2_M \sum_{i,j} |\phi_{Bi}\phi_{\widetilde{B}}^j
-({\det}'\phi_M)_i^j|^2\nonumber\\
&+&{\lambda_B^2}\mbox{Tr}(|\phi_{\widetilde{B}}\phi_M|^2
+|\phi_B\phi_M|^2) \nonumber \\
&+& m_B^2\mbox{Tr}|\phi_B|^2 
+ m_{\widetilde{B}}^2 \mbox{Tr} |\phi_{\widetilde{B}}|^2
m_M^2\mbox{Tr}|\phi_M|^2  \nonumber \\ 
&-& (h'\phi_{Bi}\phi^i_{Mj}
\phi_{\widetilde{B}}^j+ h.c.)
\end{eqnarray}
where $\lambda_M= 1/(k_M\Lambda^{2N_c-1})$, 
$\lambda_B= 1/(k_B\Lambda^{2N_c-1})$, and 
$(\det'M)_i^j \equiv \partial \det M/\partial M^i_j$.
The minimum of superpotential has been analyzed along the 
diagonal direction,
$M^i_j=M_i\delta^i_j$. 
Choosing a special direction with
$B_i\widetilde{B}^j -(\mbox{det}'M)_i^j=0$,
considering  a simple case  where 
$B_i=\widetilde{B}^i=B$, $M_{(i)}=M$,
and assuming $B$ and $M$ are real,
we write the scalar potential as
\begin{eqnarray}
{V \over N_c+1}&=& 2\lambda_B^2\phi_M^{N_c+2}-2h'\phi_M^{N_c+1}\nonumber\\ 
&+& (m_B^2+m_{\widetilde{B}}^2)\phi_M^{N_c}+m_M^2\phi_M^2
\end{eqnarray}
This scalar potential yields that
if $h'$ is sufficiently large compared with soft scalar masses, 
there will arise vacua with $\langle M\rangle \neq 0$, 
$\langle B\rangle \neq 0$,$\langle\widetilde{B}\rangle \neq 0$,
thus we get the vacua with chiral symmetry breaking but
baryon number symmetry also spontaneously breaking. 

When $N_f{\leq}N_c$, there is no possible trilinear term, but
a new quadratic SUSY breaking term,
$h_BB\widetilde{B}+h.c.$
can be introduced.
 However, it produces no new effects
compared with the case with only quadratic terms.


In summary, we studied the softly broken $N=1$ supersymmetric QCD with
the inclusion of the trilinear terms and find that in comparison
with only the quadratic soft breaking term, some
remarkable effects on the phase structure can be produced.
First, In case $N_f>N_c+1$, depending on trilinear coupling
constant and other soft breaking parameters, we get vacua 
with both the chiral symmetry breaking
and baryon number violation. Furthermore, with various choices of soft
breaking parameters, we find different phase structures. Whereas
in case with only quadratic breaking term, only exotic vacua with unbroken
chiral symmetry arise. In addition, Seiberg's duality seems to persist
in both chiral symmetry broken and unbroken phases. Second
when $N_f=N_c+1$, if the trilinear coupling is strong enough, 
there will arise vacua with spontaneously chiral symmetry breaking but 
baryon number violation. In the case with only quadratic terms, the origin
of the moduli space is the only vacuum and is hence
$SU(N_f)_L{\times}SU(N_f)_R{\times}U(1)_B{\times}U(1)_R$ i
nvariant. The significance of
this investigation is providing an enlightenment that there is a long way to
go for us to understand the non-perturbative QCD starting from  
the soft breaking of supersymmetric QCD.


\end{multicols}

\end{document}